\documentclass[namedreferences,hyperref,optionalrh]{spr-sola}
\usepackage{graphicx}        
\usepackage[export]{adjustbox}  
\usepackage{hyperref} 
\usepackage{color}           

\chardef\us=`\_

\begin{document}

\begin{frontmatter}
  
\title{Two-Part Interplanetary Type II Solar Radio Bursts}%

\author[addressref=aff1,corref,email={silpoh@utu.fi}]{\inits{S.}\fnm{Silja}~\snm{Pohjolainen}\orcid{0000-0001-7175-9370}}
\address[id=aff1]{Tuorla Observatory, Department of Physics and Astronomy,
University of Turku, Turku, Finland}

\runningauthor{S. Pohjolainen}
\runningtitle{Two-Part Interplanetary Type II Solar Radio Bursts}

\begin{abstract}
Two similar-looking, two-part interplanetary type II burst events from 2003
and 2012 are reported and analysed. The 2012 event was observed from three
different viewing angles, enabling comparisons between the spacecraft data.
In these two events, a diffuse wide-band type II radio burst was followed by
a type II burst that showed emission at the fundamental and harmonic (F-H)
plasma frequencies, and these emission bands were also slightly curved in
their frequency-time evolution. Both events were associated with high-speed,
halo-type coronal mass ejections (CMEs). In both events, the diffuse
type II burst was most probably created by a bow shock at the leading front
of the CME. However, for the later-appearing F-H type II burst there are at
least two possible explanations. In the 2003 event there is evidence of CME
interaction with a streamer, with a possible shift from a bow shock to a
CME flank shock. In the 2012 event a separate white-light shock front was
observed at lower heights, and it could have acted as the driver of the F-H
type II burst. There is also some speculation on the existence of two
separate CMEs, launched from the same active region, close in time.  
The reason for the diffuse type II burst being visible only from one viewing
direction (STEREO-A), and the ending of the diffuse emission before the
F-H type II burst appears, still need explanations. 
\end{abstract}


\end{frontmatter}


\section{Introduction}

Solar radio bursts of type II appear as frequency-drifting emission lanes
in radio dynamic spectra, with a drift from high to low frequencies. They can
be observed at decimeter--meter wavelengths \citep{kumari2023} with
ground-based telescopes, and from decameter to kilometer waves
\citep{gopal2019} with space instrumentation. Space observations at
frequencies lower than 15 MHz are often called interplanetary.

The frequency drift is generally thought to imply a propagating shock wave
in the solar atmosphere, where the decreasing plasma frequency $f_p$ (in Hz),
$f_p = 9000 \sqrt{n_e}$, is caused by the decreasing electron density
$n_e$ (in cm$^{-3}$) along the transient or shock driver path
\citep{dulk85}. Plasma emission is generated by escaping radiation
near the local plasma frequency or it's harmonics, in a multistage process
that include Langmuir turbulence and its non-linear evolution
\citep{melrose87,mann2022}. Type II bursts can often be identified from the
fundamental-harmonic (F-H) emission lane pairs, but harmonic emission is
not always observed. At meter wavelengths, harmonic emission is typically
stronger in intensity than fundamental, but at longer wavelengths the opposite
can be observed. At kilometer waves, type II bursts are characterized
by the absence of clear harmonic structure \citep{chernov2021,cane2005}. 

A distinct class of type II bursts that shows single-lane, diffuse emission
with a wide frequency band was reported by \cite{bastian2007}.
\cite{cane2005} had already noted that single-lane, broad-band interplanetary
(IP) type II bursts are common in connection with fast and large CMEs.
\cite{bastian2007} suggested that the emission may not be due to plasma
radiation at all, but instead it could be due to incoherent synchrotron
radiation from near-relativistic electrons that are trapped within a
fast-moving coronal mass ejection (CME) or sheath. The directionality of
emission could be significant for plasma radiation, but it is not expected
for synchrotron emission if the electron distribution function is nearly
isotropic. Synchrotron emission would also fade away and cease in a
certain time frame, as the magnetic field strength is decreasing in the
IP space. In case of long-duration events and if the synchrotron-emitting
electrons are shock-accelerated, the electron energies would need to
increase constantly, to compensate for the magnetic field.

After the first speculation on the emission mechanism, wide-band type II
bursts were studied further, for example with a statistical sample
from 2001--2011 \citep{pohjolainen2013}.
Almost all of the analyzed 25 wide-band type II bursts were associated with
very high-speed CMEs that originated from different parts of the solar
hemisphere, i.e., no directional effects were found.
From the data set, 18 bursts were estimated to have radio source heights
that corresponded to the CME leading front heights, suggesting that they
were created by bow shocks ahead of the CMEs. It was further suggested that
the wide-band and diffuse type II bursts were formed when specific CME
leading edge structures and/or special shock conditions were present.  

In the \cite{bastian2007} event, the diffuse single-lane emission
(named type II-S) was followed by a F-H emission lane pair (named type II-P),
dividing the event into two parts. At least the later part
of the burst thus indicated plasma radiation. The emission of the lane pair
was also slightly curved in frequency, i.e., the frequency drift made a
turn to higher frequencies, instead of drifting steadily toward lower
frequencies. 

Curved and fragmented structures within decimetric--metric type II bursts
have been reported by, e.g., \cite{pohjolainen2008}. In this particular case,
numerical MHD simulations showed that this kind of type II burst emission
can be created when a coronal shock, driven by a CME, passes through a
system of dense loops. As the radio emission reflects the varying loop
densities and the densities in between the loops (atmospheric background),
the result is drifts and shifts within the F-H emission lanes. The speed
of the shock driver determines the overall frequency drift of the burst,
but the effects of the changing densities are superposed on this emission.
In IP type II bursts, shifts and jumps to higher frequencies have also been
observed \citep{al-hamadani2017}. One plausible explanation is CME shock
interaction with a streamer \citep{feng2012}, but basically any coronal density
variation along the propagation path will affect the shape of the radio
emission in the dynamic spectrum. 

In addition to radio type II bursts, also radio type IV bursts are associated
with CMEs. At decimeter--meter wavelengths they imply the lift-off stage
of a CME, with emitting electrons trapped within magnetic structures.
At longer wavelengths, the outward-moving plasmoid can create continuum
emission that consists of both synchrotron emission from the trapped
electrons and plasma emission created by the propagating transient
\citep{hillaris2016,mohan2024}. Large-scale EUV waves \citep{mann2023}
and dimmings \citep{jin2022} have been associated with CMEs and radio
type II and type IV bursts. 

\begin{figure}
  \centering
  \includegraphics[width=\textwidth,valign=t]{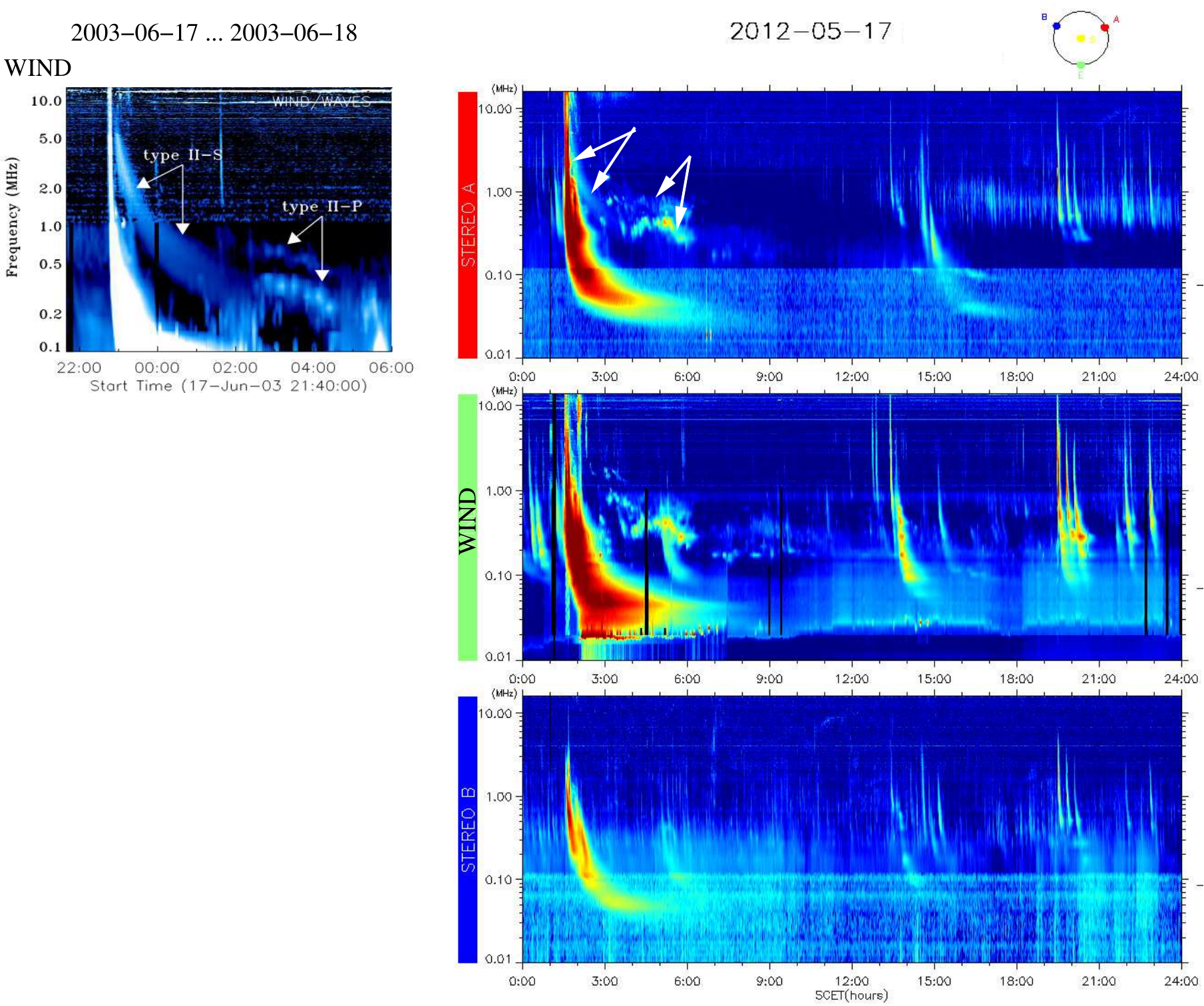}
  \caption{Observations of two type II radio burst events,
    on 17-18 June 2003 and on 17~May 2012, that show a diffuse wide-band
    burst followed by a fundamental-harmonic (F-H) emission lane pair.
    The dynamic spectrum from the  2003 event is from \cite{bastian2007},
    where the wide-band part (type II-S) and the F-H part (type II-P) are 
    indicated with arrows. The 2012 event could be observed from three
    different viewing angles, and the spacecraft positions are shown 
    in the drawing (top right corner in the plot). The similar type II
    burst structures are indicated with arrows in the STEREO-A spectrum.
    }
  \label{fig:bursts}
\end{figure}

The two-part burst on 17-18 June 2003 that was analysed by \cite{bastian2007}
could have been a single one event, if no other similar events are observed. 
However, an almost identical-looking event occurred on 17 May 2012.
The similarity of these events is most obvious when looking at the
STEREO-A/SWAVES dynamic spectrum on the 2012 event and comparing it with the
{\it Wind}/WAVES spectrum on the 2003 event, see Figure \ref{fig:bursts}.
For the 2012 event, it is possible to analyse radio observations from three
different viewing angles using STEREO-A, STEREO-B, and {\it Wind} radio
data, and compare their data with white-light and extreme ultraviolet
(EUV) imaging observations. This study aims to find out why these events
look so similar, and what could be the reason for the two-part radio
emission structures.

Observing instruments and data are described in Section \ref{data}.
Observations of flares, CMEs, energetic particles, and shock arrivals 
are summarised in Section \ref{flares-cmes}. Section \ref{radioemission}
describes the radio emission features, explains how radio source heights are
determined, and compare them with imaging observations at other
wavelengths. In Section \ref{conclusions} we summarise our findings and
discuss their relevance. In Section \ref{discussion} we consider other
issues related to this research topic. 

\section{Observations and Data}
\label{data}

For the 2003 radio event, only one viewing direction was available from
the L1 location along the Sun--Earth line. For the 2012 event, radio
dynamic spectra were taken from three different locations from three
spacecraft orbiting the Sun, and we could compare the radio spectral
characteristics with EUV and white-light images obtained from these
same locations.  

From the location of L1, {\it Solar and Heliospheric Observatory} (SOHO)
spacecraft \citep{domingo95} provides extreme ultraviolet (EUV) imaging
of the solar disc with the {\it Extreme ultraviolet Imaging Telescope} (EIT).
SOHO also provides coronagraph images taken with the
{\it Large Angle and Spectrometric Coronagraph} (LASCO) C2 and C3 instruments,
up to 32 R$_{\odot}$ distances. 
The {\it Solar Terrestrial Relations Observatory Ahead and Behind}
(STEREO A and B) twin spacecraft \citep{kaiser2008} carry five {\it Sun Earth
Connection Coronal and Heliospheric Investigations} (SECCHI)
imagers. The SECCHI imagers include {\it Extreme Ultraviolet Imager} (EUVI)
that provides EUV images of the solar disc, two coronagraphs COR1 and COR2
that provide Sun-centered images up to a distance of 15 R$_{\odot}$, and
two heliocentric imagers HI1 and HI2 that provide white light images from
the outer part of COR2 field of view, up to a distance of Earth orbit.  

Images and associated data products were obtained from the
CDAW LASCO CME Catalog at  \url{cdaw.gsfc.nasa.gov} and by using
the data and software in \textsf{JHelioviewer} \citep{muller2017}.

Radio dynamic spectra at decimeter-meter wavelengths can be obtained from
ground-based observatories. At decameter--hectometer (DH) wavelengths
dynamic spectra can be obtained from {\it Wind}/WAVES \citep{bougeret95}
located in L1, close to SOHO with the same field of view.
STEREO/SWAVES instrumentation \citep{bougeret2008} in the two STEREO
spacecraft are very similar to the {\it Wind}/WAVES instruments.  

\begin{figure}
  \centering
  \includegraphics[width=0.49\textwidth]{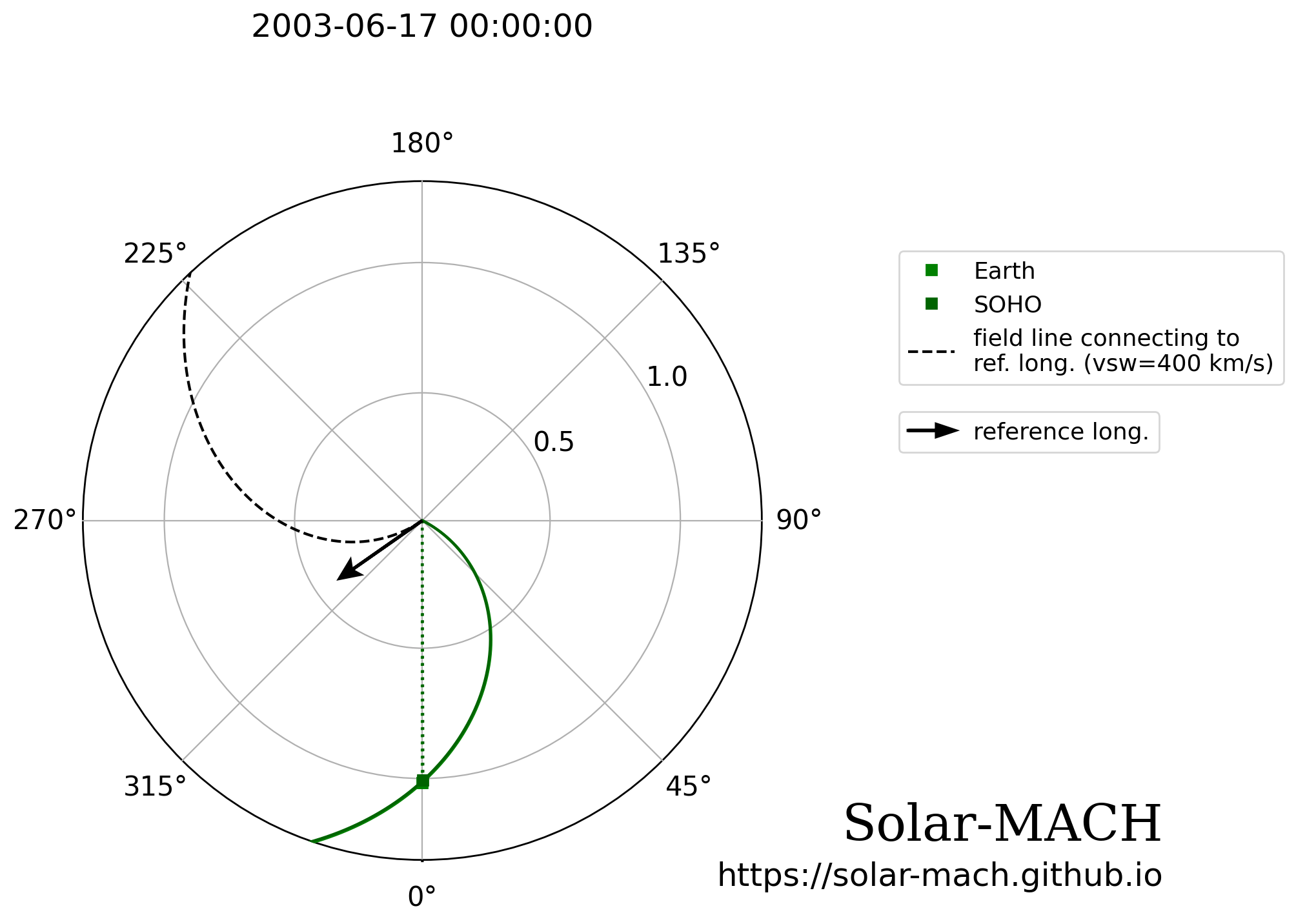}
  \includegraphics[width=0.49\textwidth]{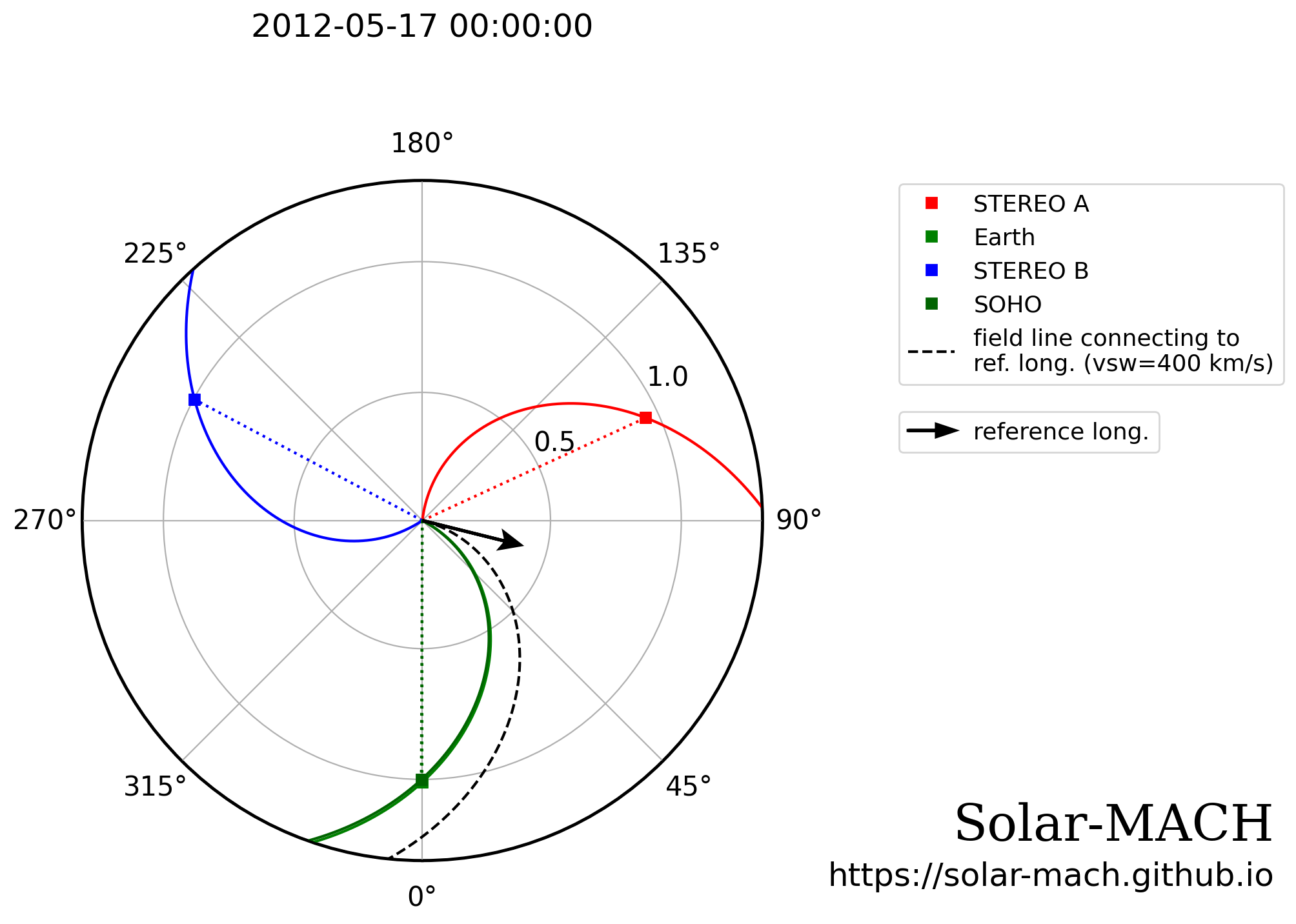}
  \caption{\textsf{Solar-MACH} plots for 17 June 2003 (left) and 17 May 2012
    (right). Black arrows indicate the flare longitudes and black dashed
    lines their magnetic connectivities.  
    }
  \label{fig:MACH}
\end{figure}

For the 2003 event only {\it Wind}/WAVES observations are available,
from Earth view. In 2012, also STEREO-A and STEREO-B made radio observations
from two different positions. The separation angles between Earth and
the spacecraft were 114.8$^{\circ}$ (Earth\,--\,A) and
117.6$^{\circ}$ (Earth\,--\,B), and the angle between A and B was
127.5$^{\circ}$. The spacecraft positions at the event times are indicated in
Figure~\ref{fig:MACH}. The figure also shows flare locations in the
two events, with the available observing directions. 

\section{Event Descriptions}
\label{flares-cmes}

\subsection{Flares}

\begin{figure}[!ht]
  \centering
  \includegraphics[width=0.42\textwidth]{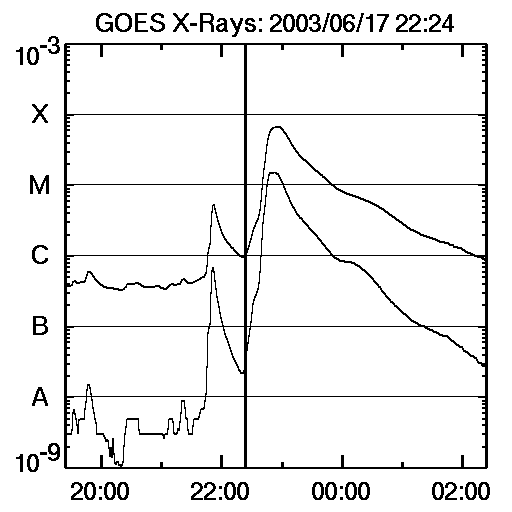}
  \includegraphics[width=0.4\textwidth]{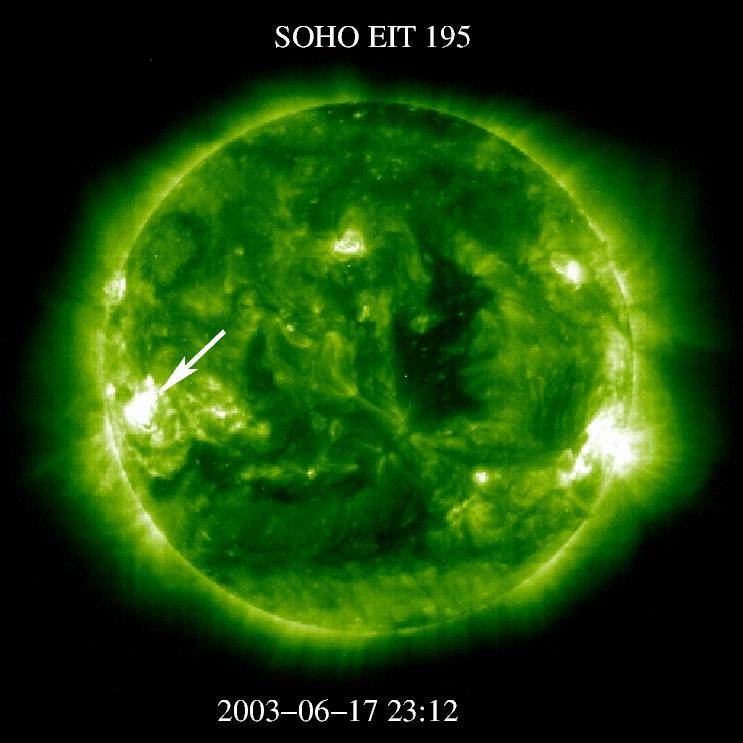}
  \includegraphics[width=0.42\textwidth]{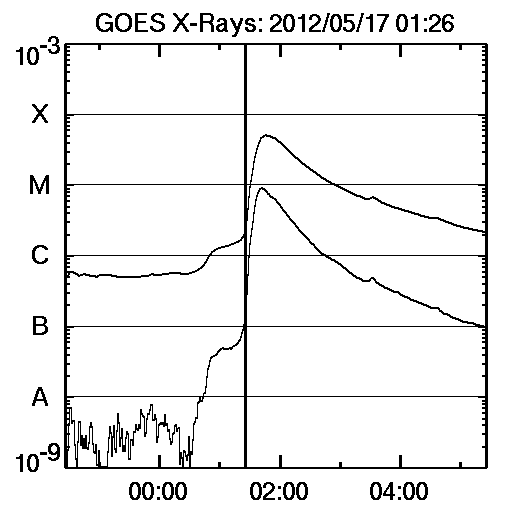}
  \includegraphics[width=0.4\textwidth]{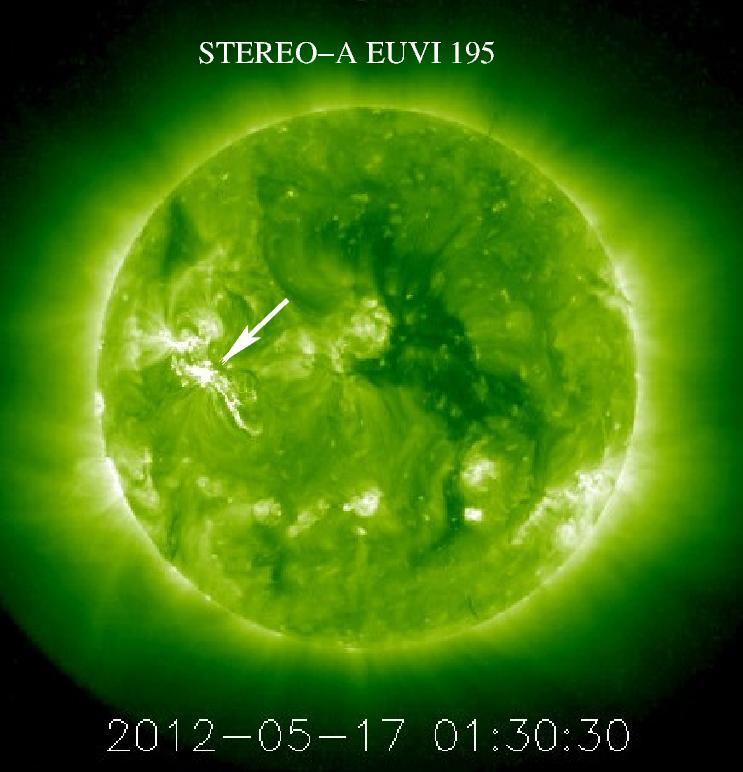}
  \caption{Flare intensities in soft X-rays observed by GOES (Earth view).
    The 2003 flare was located at S07E55 and the 2012 flare at N11W76.
    In STEREO-A view the 2012 flare occurred near N11E42.
    The flare locations are indicated with arrows in the EUV images, 
    observed by SOHO/EIT and STEREO-A/EUVI at 195 \AA \, wavelength. 
    }
  \label{fig:GOES}
\end{figure}

The origin of the 17-18 June 2003 event was a flare located at
S07\,E55 in AR~10386. The flare was listed to start at 22:27 UT, and
peak at 22:55 UT. The flare intensity was GOES class M6.8. 
The flare was preceded by a GOES C5.2 flare in the same region, 
which was listed to start at 21:44 UT and peak at 21:53 UT.
The GOES plot in Figure \ref{fig:GOES} shows that in X-rays the
M6.8 flare starts in the intensity fall phase of the C5.2 flare.
The SOHO/EIT 195 \AA \, image shows the flare location at S07E55
(Earth view). No clear EUV wave was observed. 

The 17 May 2012 event was associated with a flare located at N11\,W76
in AR 11476. In STEREO-A view the flare was located approximately at
N11\,E42. The flare was listed to start at 01:25 UT, and peak at 01:47 UT.
The flare had GOES class M5.1 intensity (Earth view). 
The STEREO-A/EUVI 195 \AA \, image in Figure~\ref{fig:GOES} shows the
flare location near N11E42. The flare was associated with an EUV wave
that was best observed from STEREO-A.

\subsection{CMEs}
\label{cmes}

Both events were associated with fast halo-type CMEs.
On 17-18 June 2003 the halo CME speed, from a linear fit to the leading front
listed in the LASCO CME Catalog, was 1812~km~s$^{-1}$ and the speed remained
approximately the same during the observations.

\begin{figure}
  \centering
  \includegraphics[width=0.3\textwidth]{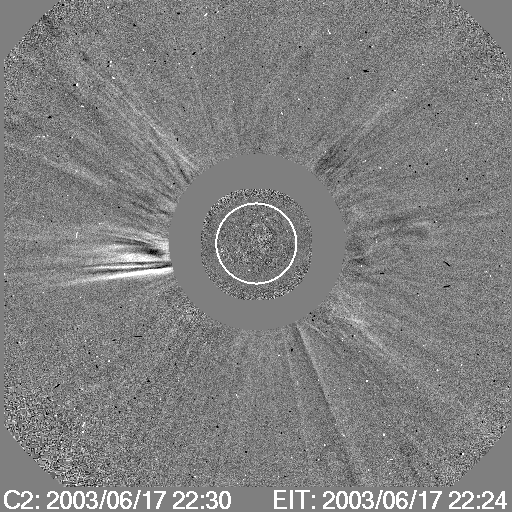}
  \includegraphics[width=0.3\textwidth]{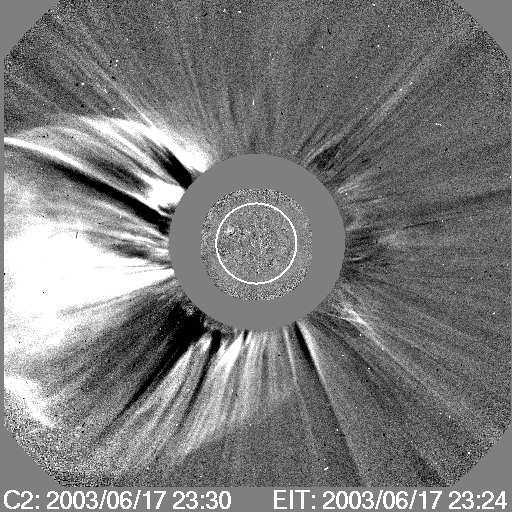}
  \includegraphics[width=0.315\textwidth]{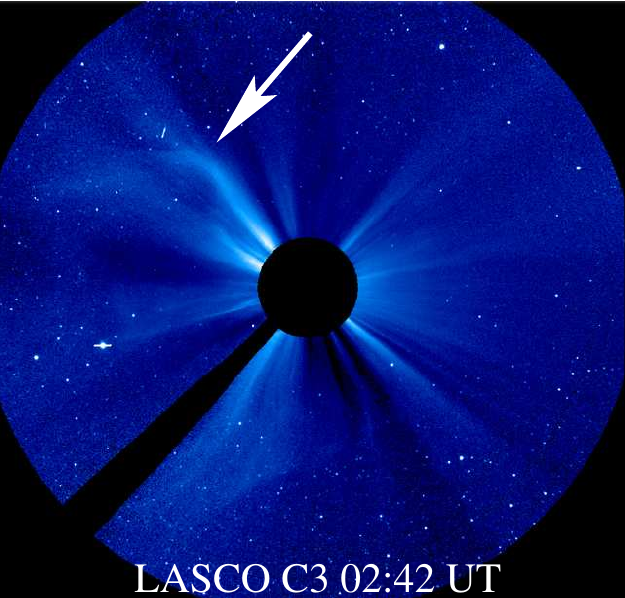}
  \caption{SOHO/LASCO-C2 difference images at 22:30 and 23:30~UT
    on 17 June 2003. The narrow-width eastward CME structure at 22:30 UT
    originated from the C5.2 flare. At 23:30~UT the front of the halo CME,
    associated with the M6.8 flare, had already passed the LASCO-C2
    field of view. At 02:42~UT the LASCO-C3 image shows a bending of
    a streamer (location indicated with an arrow). 
        }
  \label{fig:lascoC2-2003}
\end{figure}

The CME leading front was first observed by SOHO/LASCO-C3 on 17 June 2003
at 23:18 UT, at height 7.06 R$_{\odot}$. This late first-observed height
was due to a gap in C2 observations. Figure~\ref{fig:lascoC2-2003} shows
two LASCO-C2 difference images, at 22:30 UT and 23:30~UT. At 22:30 UT
a narrow-width bright structure preceded the halo CME, but it was not
listed as a separate CME. At 23:30~UT the halo CME front had already
passed the C2-instrument field of view.

Several streamers are visible in the coronagraph images. The streamer
on the north-eastern side of the CME looks to be affected from
02:42~UT onward on 18 June 2003.

On 17 May 2012 the linear fit CME speed was 1582~km~s$^{-1}$.
The halo CME was decelerating, with a last-observed speed of
1320~km~s$^{-1}$ at height 26 R$_{\odot}$.

The CME front was first observed by SOHO/LASCO on 17 May 2012 at 01:48~UT,
at height 3.61 R$_{\odot}$, but STEREO-A/COR1 observed the CME front already
at 01:40 UT, at height 2.05 R$_{\odot}$. The observed CME heights from
STEREO-A were, and also remained, lower than the LASCO CME heights.
The CME heights observed from STEREO-B were even lower. From the height
differences and spacecraft positions we can conclude that the CME was
propagating to a direction in between Earth/SOHO and STEREO-A (both
observed the leading front projected on the sky). STEREO-B did not observe
the actual front as it was moving away, on the opposite side of the Sun. 
Streamers were also present on both sides of the expanding CME.

\begin{figure}[!hb]
  \centering
  \includegraphics[width=\textwidth]{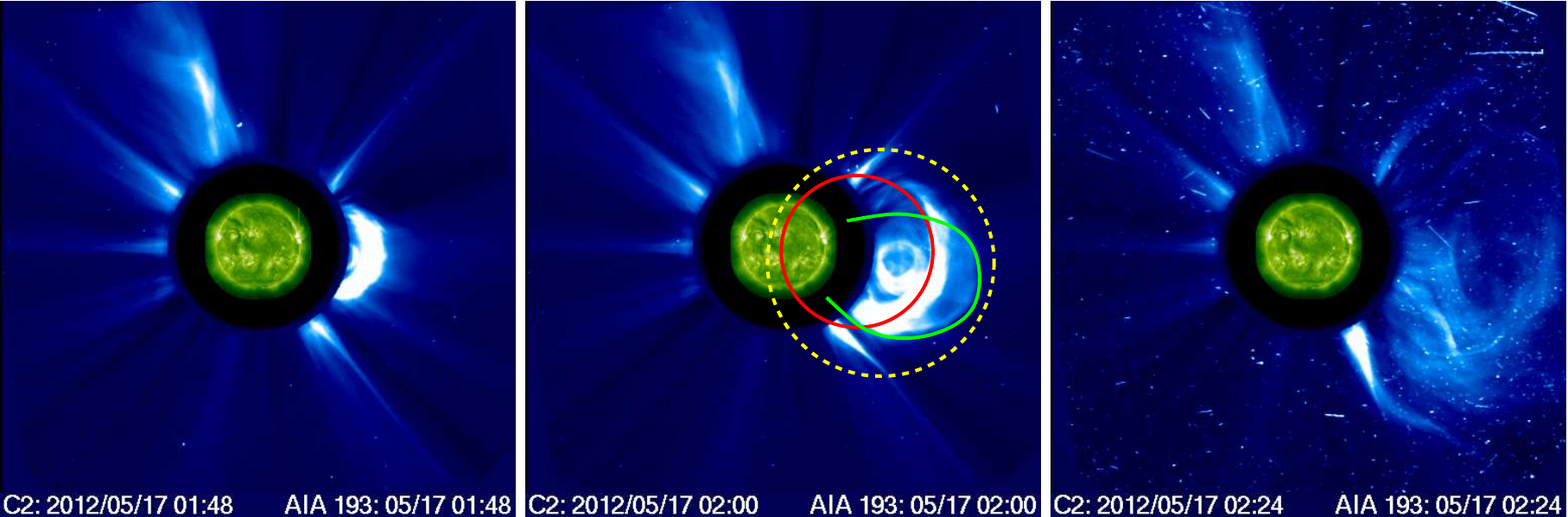}
  \caption{CME event on 17 May 2012, from SOHO/LASCO-C2 observations at
    01:48 UT, 02:00 UT, and 02:24 UT.
    In the 02:00 UT image the two CMEs suggested by \cite{shen2013} are
    outlined roughly with red (CME1) and green (CME2) lines. The yellow
    dashed line approximates the location of the wave-halo suggested
    by \cite{grechnev2024}. The most-affected streamer is located
    on the south-western side of the CME.
    }
  \label{fig:lascoC2-2012}
\end{figure}

The eruptions in this 2012 solar event have been studied by \cite{shen2013}
and more recently by \cite{grechnev2024}. \cite{shen2013} suggested
that there were two eruptions that originated from the same active region
close in time, within minutes, and they formed two separate CMEs.
They used the Graduated Cylindrical Shell (GCS) model to study how
the two CMEs propagated. CME1 was outlined as a halo (indicated with a
red line in Figure \ref{fig:lascoC2-2012}), and CME2 was propagating
sideways in a more narrow cone (green line in Figure \ref{fig:lascoC2-2012}).
According to \cite{shen2013}, in the STEREO-A observations these
structures overlap and appear to be propagating together.
\cite{shen2013} also proposed that there were two separate shocks
caused by the two CMEs. 

\cite{grechnev2024} studied these eruptions further and they also identified
two separate eruptions, but now located behind a CME-bubble front.
In their interpretation, this led to two propagating wave-like disturbances
within one expanding CME. The two disturbances later merged into a
single, stronger, and faster shock, a wave-halo (location indicated with
a dashed yellow line in Figure~\ref{fig:lascoC2-2012}). At heights
exceeding ten solar radii, after 03:00 UT, the wave-halo front looked
to fall behind the fastest LASCO CME-body front. So, in some aspects,
this solar event still remains unclear.


\subsection{Energetic Particles and Interplanetary Transients}

Intense flares and shock waves associated with fast CMEs have been named
as the source of solar energetic particle (SEP) events, where protons
are accelerated to relativistic energies. They can also cause a signal
in ground-based detectors like neutron monitors. These are called ground
level enhancement (GLE) events. Propagating shock waves and interplanetary
CMEs (ICME) can be observed in-situ, as they change the local plasma
parameters. The database of interplanetary shocks is
at \url{https://ipshocks.helsinki.fi/}. ICME observations from
2007--2016 using the STEREO spacecraft are available at
\url{https://stereo-ssc.nascom.nasa.gov/pub/ins_data/impact/level3/},
see also \cite{jian2018}. In ICME-related shocks, the shock arrival is
followed by the magnetic cloud passage.

The 17-18 June 2003 burst source was not well-connected to Earth, as it was
located near the solar east limb. The event was not listed as a GLE event,
but it was a proton event with proton flux of 24~pfu at energies $>$10 MeV,
see the NOAA SEP event list at \url{umbra.nascom.nasa.gov/SEP/}.

{\it Wind} observed shock arrival on 20 June 2003 at 08:02 UT, with shock
speed 638~km~s$^{-1}$. The travel time from the Sun to the spacecraft would
be 57 hours, with mean speed of 725 km s$^{-1}$.

The 17 May 2012 burst source was located near the west limb, with very good
connection to Earth, see the magnetic field lines in the \textsf{Solar-MACH}
plots in Figure \ref{fig:MACH}. The event was a GLE event number 71,
the first of only two GLEs to occur in Solar Cycle 24. Several studies
have investigated the SEP acceleration in this event, using the available
multi-spacecraft in-situ measurements \citep{grechnev2024}. The proton
flux was 255~pfu at energies $>$10 MeV \citep{kocharov2018}.

STEREO-A observed shock arrival on 18 May 2012, at 12:43 UT, with speed
793 km s$^{-1}$. The travel time to the spacecraft would be 35 hours, with
mean speed of 1200 km s$^{-1}$.
{\it Wind} observed shock arrival on 20 May 2012, at 01:20~UT, with
speed 439 km s$^{-1}$.  The travel time to the spacecraft would be 72 hours,
with mean speed of 575 km s$^{-1}$.
The time separation between the arrival times at STEREO-A and {\it Wind}
was 37 hours.


\section{Radio Emission and Source Heights}
\label{radioemission}

An essential part in comparing CME heights to radio source heights is the
conversion from emission frequency to height. As plasma frequency depends
only on the electron density in the surrounding medium, radio source
heights can be calculated using atmospheric density models. The different
models are explained and compared in \cite{pohjolainen2007}, see
their Table 1. In \cite{al-hamadani2017}, height errors based on the model
selection were estimated as 2.0~R$_{\odot}$ near 1 MHz, 1.0~R$_{\odot}$ near
500 kHz, and 0.3~R$_{\odot}$ near 150 kHz. This is based on the fact that
as the frequency decreases, the height difference between the models
decreases. At 10 MHz frequency where the IP radio observations begin,
the height difference between the most-extreme density models, Saito
and 10-fold Saito, is 1.6~R$_{\odot}$ \citep{pohjolainen2007}. The hybrid
atmospheric density model by \cite{vrsnak04} is adopted in the current
study, as the model starts from the high densities close to the Sun,
and also gives the low (measured) densities near Earth.

For wide-band emission, the next decision is which part of the emission
lane should be used for height calculations. If we assume that the emission
lane is widened by plasma processes, we can then use the emission lane center.
If, however, the wide lane is produced by a shock front that extends to
a large height range, for example a bow shock along a curved CME front,
then the emission lane borders can be used. For a fundamental-harmonic (F-H)
lane pair, the F-lane should always be used in calculations.

CME propagation in the wake of earlier CMEs can change the atmospheric
densities, and hence affect the observed plasma frequencies during shock
propagation. The 2003 halo CME was preceded by a narrow-width CME that
originated from the same active region, but it was not listed as a CME
in the LASCO catalog as it had no clear moving front. The flow of the
CME material is visible as a brightening in the coronagraph difference
image at 22:30~UT, shown in Figure \ref{fig:lascoC2-2003}.
The 2012 halo CME was preceded by other CMEs close in time, but
interaction with them does not seem likely. One of them propagated up to
20~R$_{\odot}$, but the CME was directed northward and had narrow angular
width of about 80 degrees. The two other CMEs were classified as poor
events, that disappeared from the images near 10~R$_{\odot}$ height. They
were directed southward and westward, and had widths of 51 and 16 degrees,
respectively. Plasma densities could have been affected by these structures,
but not in a large scale.

\begin{figure}
  \centering
  \includegraphics[width=\textwidth]{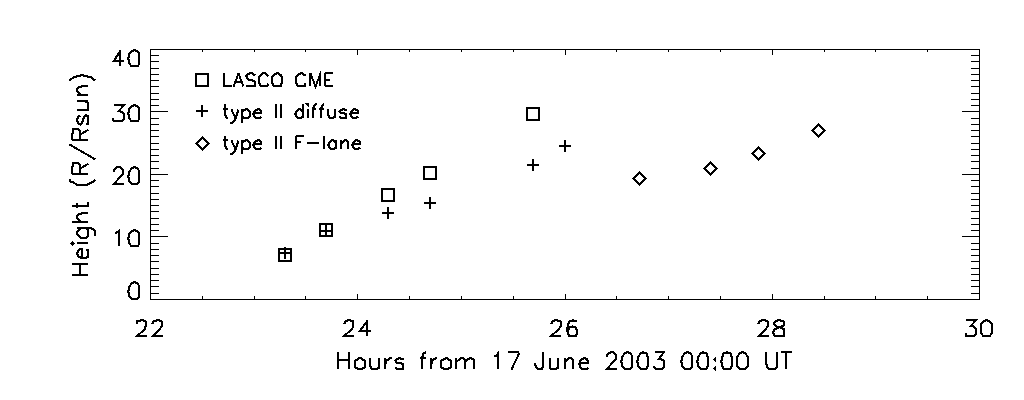}
 \includegraphics[width=\textwidth]{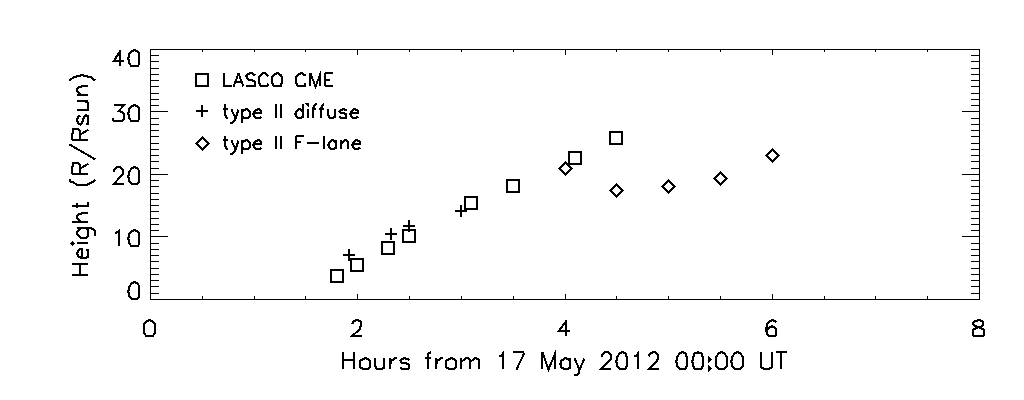}
 \caption{Height-times for the LASCO CME front (squares),
   diffuse wide-band emission lane center (crosses), and F-lane of 
   the second part of the type II burst (diamonds),
   for the 17-18 June 2003 event ({\it top}) and for the 17 May 2012
   event  ({\it bottom}). Both plots show an eight-hour interval.
   The symbol size is 2~R$_{\odot}$, which represents the maximum error
   in radio source heights due to the use of atmospheric density models,
   see text for details. 
   }
  \label{fig:heights-2003-2012}
\end{figure}

The radio source heights calculated from the diffuse wide-band emission
lane center and from the F-lane of the later, F-H part of the type II
burst in the 2003 and 2012 events are shown in
Figure \ref{fig:heights-2003-2012}. These height-time plots include
the CME leading front heights observed by SOHO/LASCO, and listed in the
LASCO CME Catalog. For the 2003 event, the radio source heights are
from {\it Wind}/WAVES observations (same viewing angle as for SOHO).
For the 2012 event, the radio source heights are from STEREO-A/SWAVES
observations, as Wind/WAVES did not observe the diffuse wide-band
type II burst at all. 


\subsection{The 17-18 June 2003 Radio Event}

\begin{figure}
  \centering
  \includegraphics[width=0.45\textwidth]{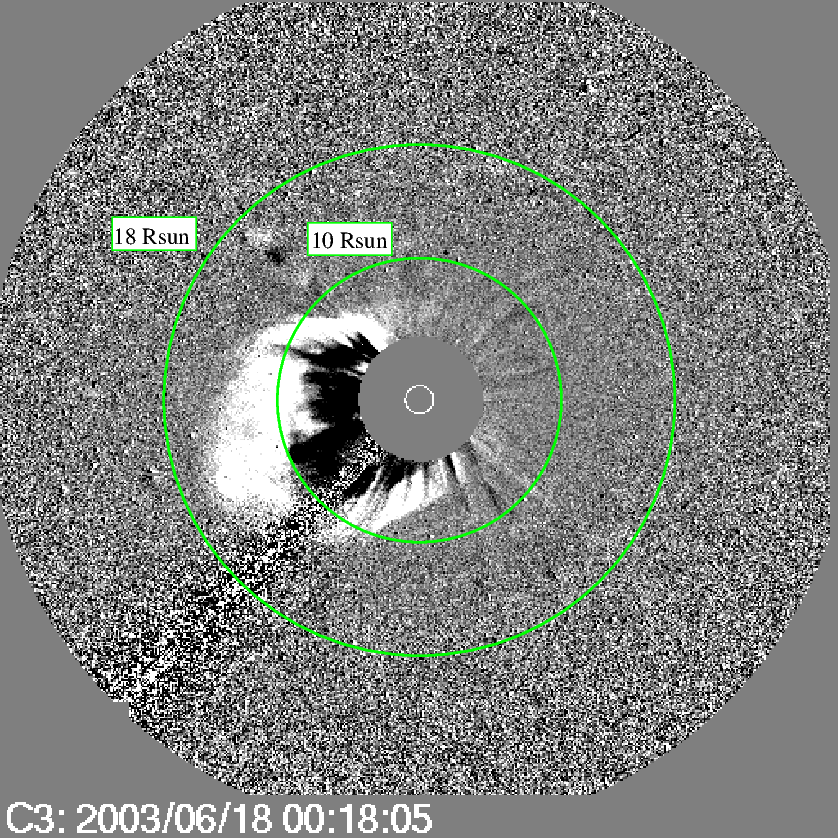}
  \includegraphics[width=0.45\textwidth]{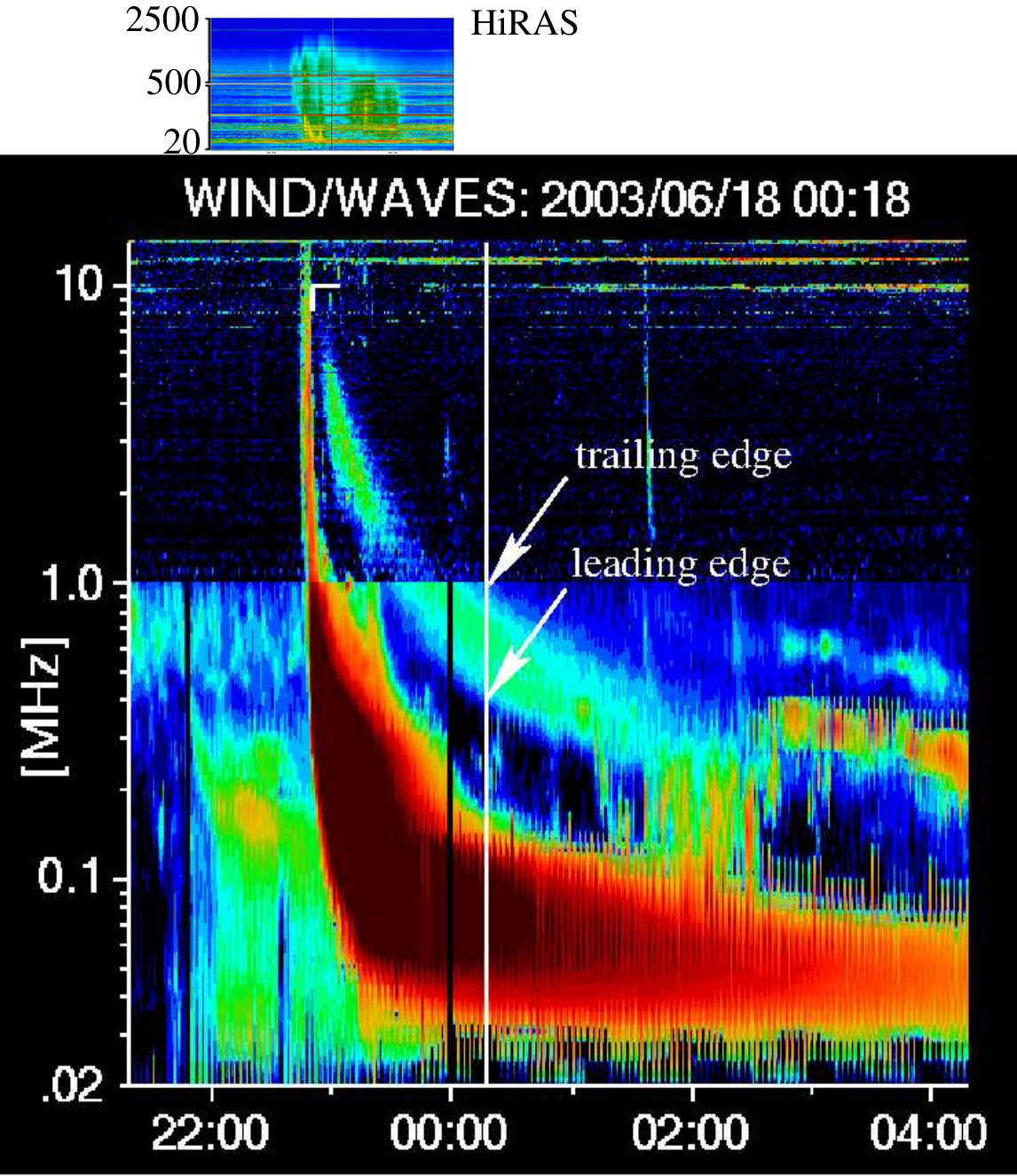}
  \caption{SOHO/LASCO-C3 difference image on 18 June 2003 at 00:18 UT.
    Green circles at heights 18~R$_{\odot}$ and 10~R$_{\odot}$ indicate
    the heights of the wide-band type II emission lane borders (leading
    and trailing edges of emission, respectively) at that time.
    The CME front covers these heights quite well. HiRAS dynamic spectrum
    (on top) shows a metric type II burst and a type IV continuum. These
    emissions did not continue to the {\it Wind}/WAVES frequencies.
   }
  \label{fig:CME-heights-2003}
\end{figure}

The radio event started with a metric type II burst at 22:50 UT, near 50 MHz,
on 17~June 2003, see the HiRAS spectrum in Figure \ref{fig:CME-heights-2003}
and NOAA spectral listings from Gulgoora and HiRAS. A metric type IV
continuum appeared at the same time, and it continued until 23:50 UT.
Both metric emissions could be recorded down to 20~MHz, but they did not
continue to the {\it Wind}/WAVES frequencies.

The diffuse wide-band type II emission lane appeared in the dynamic
spectrum at 23:00 UT, at 6~MHz center frequency. 
The heliocentric source height was $\approx$\,4.1~R$_{\odot}$,
calculated from 6 MHz using the hybrid atmospheric density model.
When the  CME was first observed at 23:18~UT, the CME leading front was
at height 7.06~R$_{\odot}$, and the diffuse wide-band type II burst at
height 7.3~R$_{\odot}$.
The separation between the CME heights and the diffuse emission source
heights grew after 23:40~UT, and the diffuse radio source
was located $\approx$\,7~R$_{\odot}$ lower than the CME leading front 
at 01:42 UT (last observation of the CME front, at 29.6~R$_{\odot}$).

We calculated also radio burst heights by using the leading and
trailing edges of the diffuse emission lane.
Figure \ref{fig:CME-heights-2003} shows a difference image of the
CME at 00:18 UT, with the calculated height range for the
diffuse type II emission (region in between the green lines).
The leading edge, at 400 kHz, has a height of 18~R$_{\odot}$,
and the trailing edge, at 1~MHz, is at 10~R$_{\odot}$ height.
Despite the height differences and also when we take into account the
differences between atmospheric density models, the diffuse type II
burst heights follow the CME front heights quite well until 00:30 UT. 

\begin{figure}
  \centering
  \includegraphics[width=0.93\textwidth]{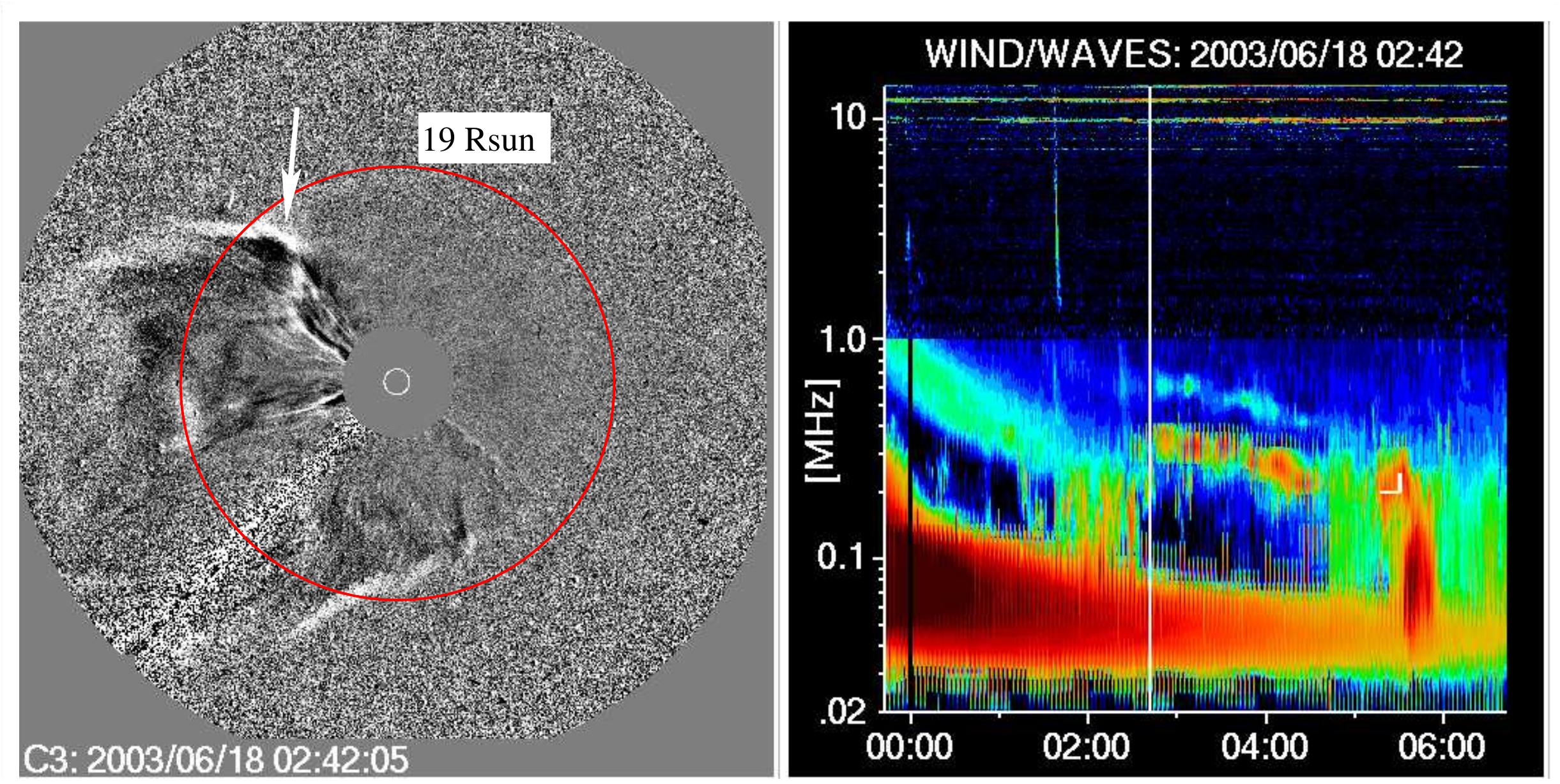}
  \caption{CME shape near the start time of the F-H emission
    lane pair on 18 June 2003, at 02:42 UT. The F-lane radio emission 
    starts at 360 kHz, which corresponds to a height of about 19 R$_{\odot}$.
    This height is indicated with a red circle in the LASCO-C3 difference
    image. The most-affected streamer region is marked with an arrow.
      }
  \label{fig:spec-comp2003}
\end{figure}

The F-H emission lane pair became visible near 02:40 UT, and after
04:30~UT the lane pair could no longer be identified among other intense
emissions (Figure~\ref{fig:spec-comp2003}). The F-lane of the burst
drifted from 360~kHz to 230~kHz, which corresponds to heliocentric
heights from 19.3~R$_{\odot}$ to 27.0~R$_{\odot}$.
The CME leading front could no longer be observed when the F-H lane
pair appeared in the dynamic spectrum, but from the extrapolated CME
heights the difference to the F-lane height at 02:43 UT was
$\approx$20~R$_{\odot}$. The CME shape near the start time of the
F-H lane emission is  shown in Figure \ref{fig:spec-comp2003},
and the red line indicates the radio F-lane source height at
19.0~R$_{\odot}$. At that height we also see a bending of a streamer,
on the north-eastern flank of the CME.  


\subsection{The 17 May 2012 Radio Event}

\begin{figure}
  \centering
  \includegraphics[width=0.9\textwidth]{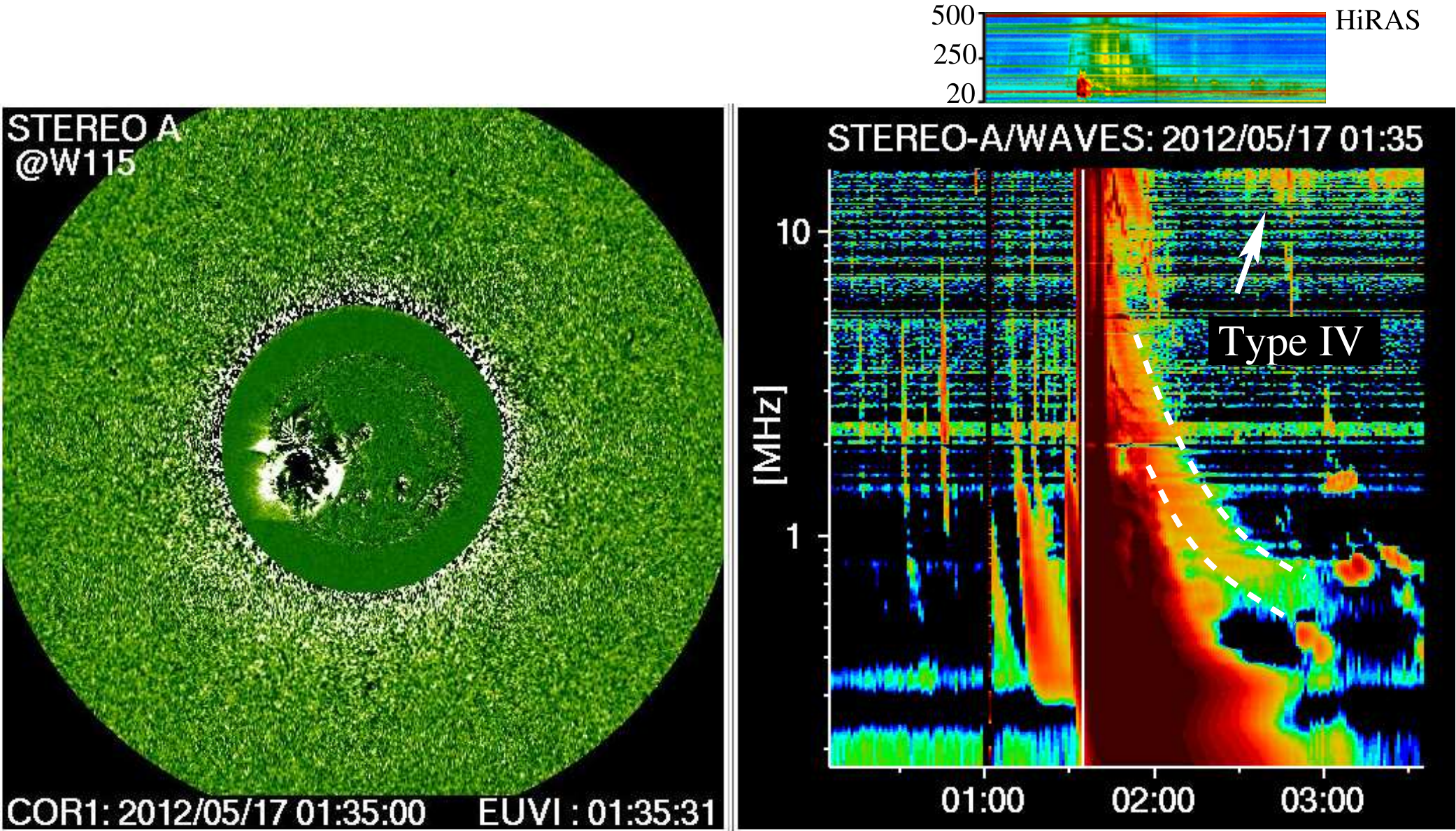}
  \includegraphics[width=0.45\textwidth]{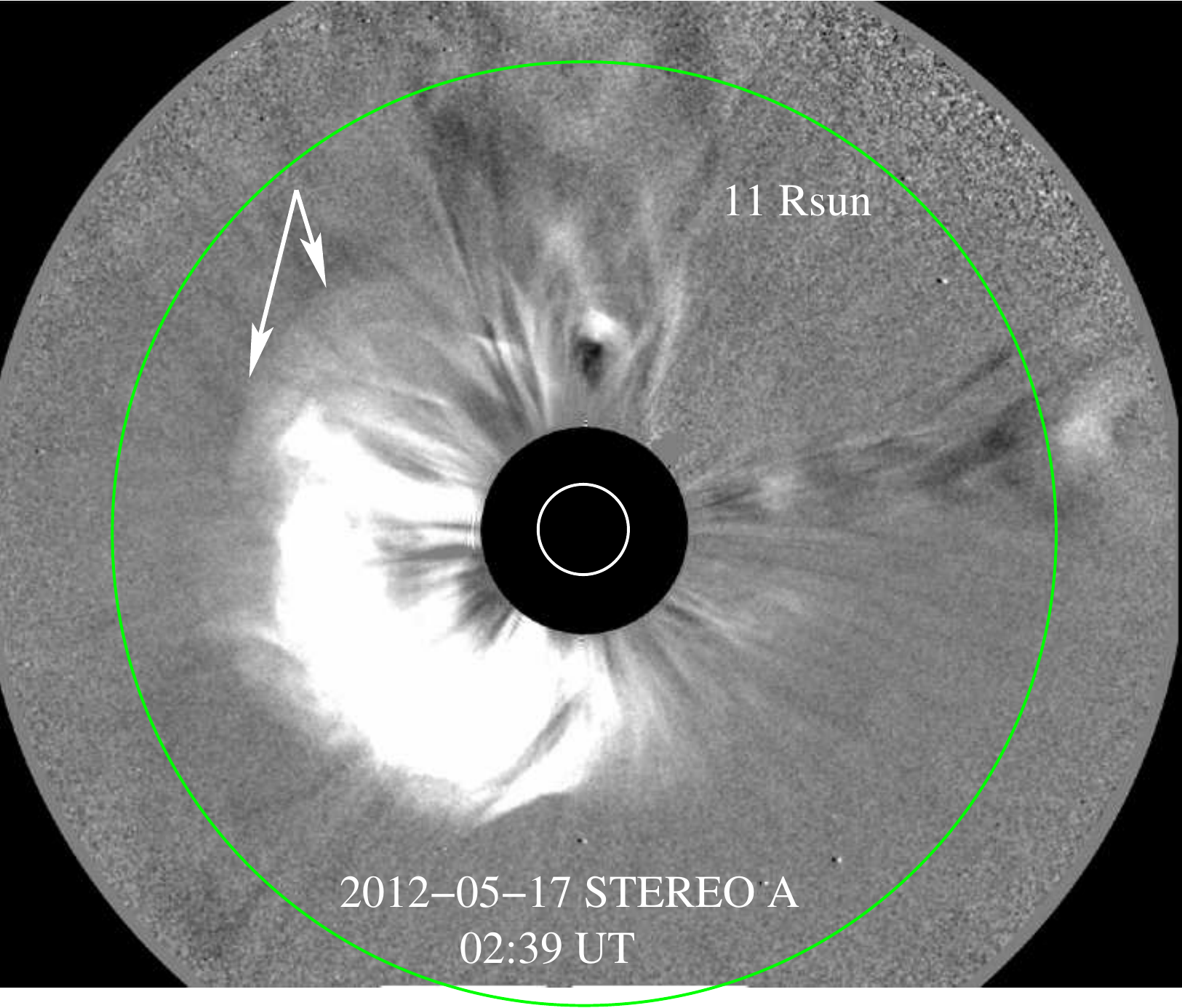}
  \includegraphics[width=0.45\textwidth]{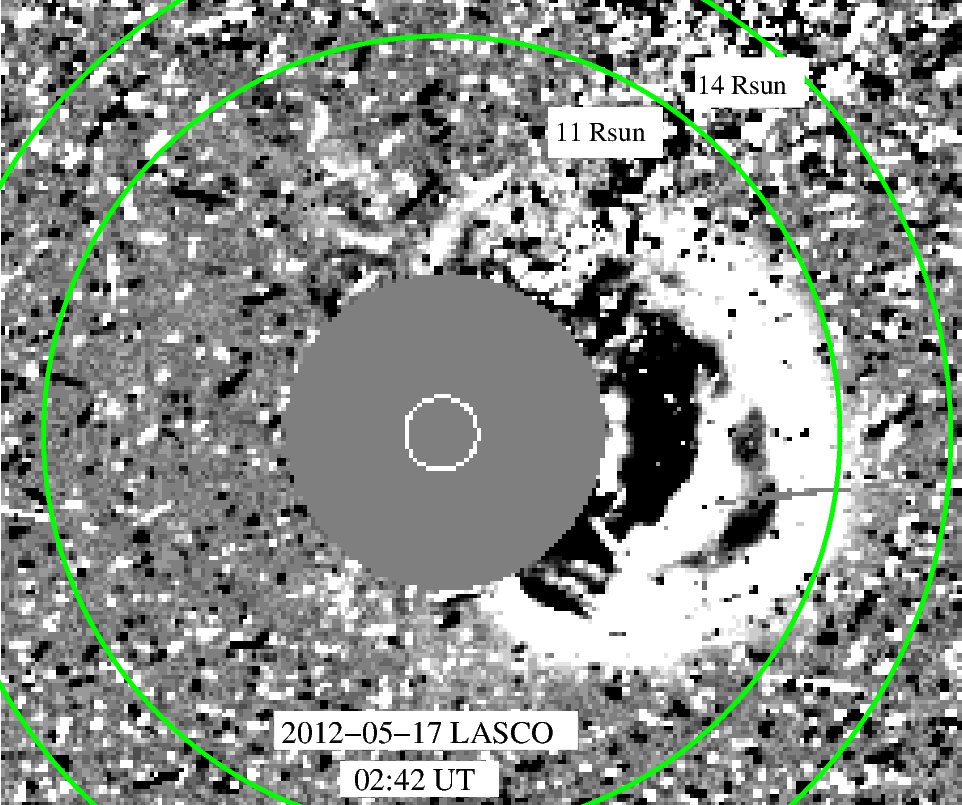}
  \caption{In the 2012 event STEREO-A/EUVI observed an EUV wave at 01:35 UT
    (top left).
    STEREO-A dynamic spectrum shows a diffuse type II burst (leading and
    trailing edges indicated with dashed lines) and a type IV burst.
    HiRAS dynamic spectrum on top shows a metric type II burst and a type
    IV continuum. 
    The CME was observed by STEREO-A/COR2 at 02:39~UT on 17 May 2012
    (bottom left) and by SOHO/LASCO-C3 at 02:42 UT (bottom right).
    The height range of the diffuse wide-band type II burst was 
    11\,--\,14~R$_{\odot}$ near those times. These heights are indicated
    with green circles on the coronagraph images. In the STEREO-A difference
    image at 02:39 UT arrows point to the white-light shock region that
    formed in the north-eastern part of the CME leading front. 
   }
  \label{fig:CME-heights-2012}
\end{figure}

In the 17 May 2012 event metric type II emission was first observed at
01:32~UT near 50 MHz, followed by a metric type IV continuum, see
HiRAS spectrum in Figure \ref{fig:CME-heights-2012}. Culgoora observations
show the metric emissions as well, and they can be viewed for example
in \cite{grechnev2024}. Both bursts look to continue to DH wavelengths,
but the type IV burst was well-observed only by STEREO-A.

\begin{figure}
  \centering
   \includegraphics[width=0.8\textwidth]{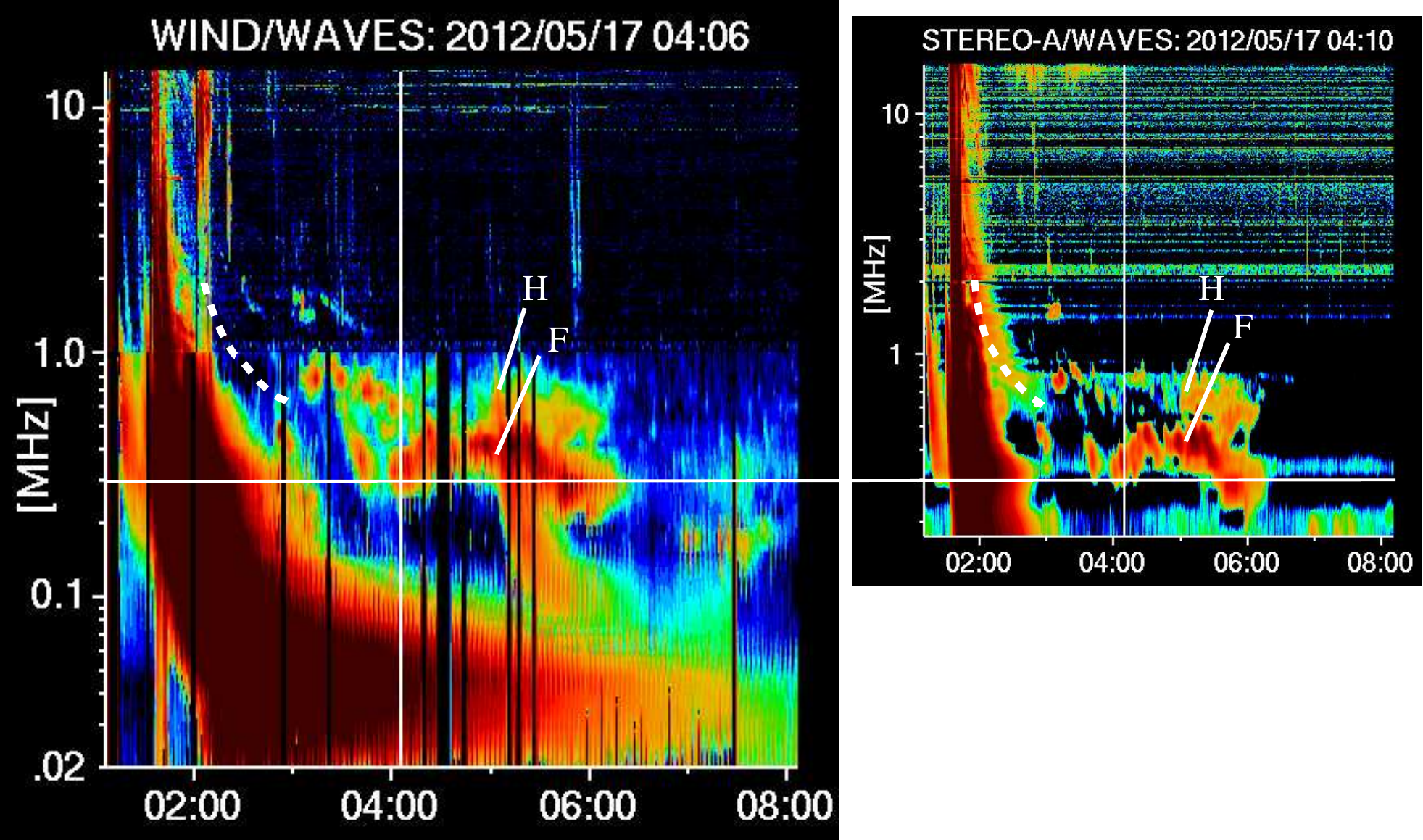}
   \caption{Comparison between the {\it Wind}/WAVES and STEREO-A/SWAVES
     dynamic spectra. Dashed white line indicates the frequencies of the
     emission lane center of the diffuse wide-band type II burst. Horizontal
     white line marks the 300 kHz frequency where the F-band first appeared.
     The F-H type II burst lanes look almost identical, but the existence
     of the lane pair is uncertain before 05:00 UT, and {\it Wind}/WAVES shows
     more structures at lower frequencies at earlier times. The diffuse
     wide-band type II burst and the type IV burst are not visible in the
     {\it Wind}/WAVES spectrum, they show up only in the STEREO-A observations.
     }
 \label{fig:wind-stereo-0410}
\end{figure}

The start time of the diffuse wide-band type II burst is uncertain, due
to the emission being mixed with strong type III bursts. The diffuse emission
lane was observed only by STEREO-A. The lane becomes visible at 01:55 UT
at 2~MHz, and the lane center corresponds to a height of 7.1 R$_{\odot}$.
When the wide-band emission disappears from the spectrum at 03:00 UT, the
lane center is at 600~kHz, and the estimated source height is 14.0~R$_{\odot}$.
Some emission fragments, at various frequencies, are observed during
03:00\,--\,04:00 UT, see also Figure \ref{fig:wind-stereo-0410}.

The leading and trailing edge heights of the diffuse band emission
are shown in Figure \ref{fig:CME-heights-2012}, indicated with green
lines in the coronagraph images. The leading edge is at height
14.0~R$_{\odot}$ at 02:40 UT, which is higher than the CME fronts observed
by SOHO/LASCO and STEREO-A.
The trailing edge of the diffuse emission, at 11.0~R$_{\odot}$, is very
near the LASCO-C3 CME front height (11.5~R$_{\odot}$), but in STEREO-A
view the CME front is at much lower height (8.6~R$_{\odot}$). We note,
again, that from the SOHO/LASCO viewing angle the diffuse emission
lane was not observed at all (white dashed line in the {\it Wind}
dynamic spectrum in Figure \ref{fig:wind-stereo-0410} shows where
the lane should have been).

\begin{figure}
  \centering
   \includegraphics[width=0.7\textwidth]{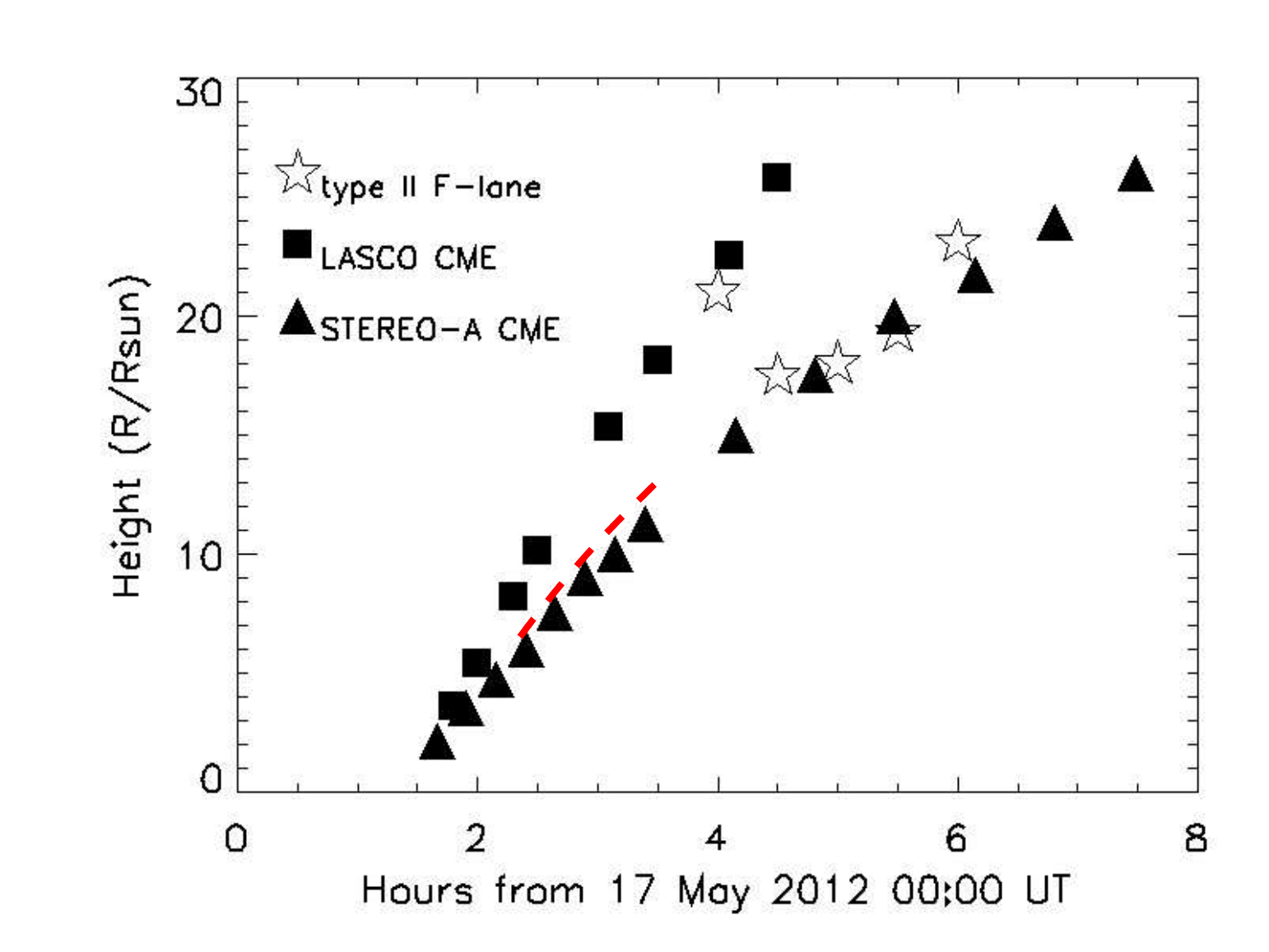}
   \caption{CME heights from SOHO/LASCO (C2 and C3) and from STEREO-A/SECCHI
     (COR1, COR2, and HI) for the 17 May 2012 event. The STEREO-A CME
     front heights are similar to the heights of the white-light shock region
     located in the north-eastern part of the CME leading front, the shock
     region heights are indicated with a red dashed line (COR2 observations).
     Estimated heights for the F-lane of the type II burst are shown
     with star symbols. The H-lane of the burst appeared only after 05:00 UT.}
 \label{fig:heights-2012-stereo}
\end{figure}

In the STEREO-A/COR2 white-light images a sharp edge of diffuse emission
is visible in the north-eastern part of the CME, indicated with
arrows in Figure~\ref{fig:CME-heights-2012}. This type of feature can
be interpreted as a density enhancement from a fast-mode MHD shock
\citep{vourlidas2003}. The white-light shock heights are indicated
with a red dashed line in Figure \ref{fig:heights-2012-stereo}, and
they are similar to the STEREO-A CME heights.
The speed of the shock front, approximately 1100~km~s$^{-1}$, was slightly
higher than the speed of the STEREO-A CME front, 1000~km~s$^{-1}$. 
The sharp-edged shock region is visible also in the STEREO-B images,
although at much lower heights, as from this viewing angle the shock
is moving away from the observer. In the SOHO/LASCO field-of-view the
white-light shock structure is located in the north-western side of
the CME, see for example Figure 4 in \cite{gopal2013}. 

STEREO-A dynamic spectrum shows a type IV burst from 02:25 UT
until 03:50 UT, at frequencies higher than 12 MHz, which looks to be a
continuation of the metric type IV emission. A type IV burst
is typically an indication of flare-accelerated electrons trapped
in post-flare loops or CME structures, and the radiation is due to
either synchrotron or plasma emission, or both, see for example
\cite{nasrin2018} and references therein.
The type IV burst disappears from the spectrum near the time when
the F-lane emission starts. Type IV bursts are known to show directivity
effects, i.e., the emission is visible only if the source region is
located within 60$^{\circ}$ from the solar disc center \citep{mohan2024}.
This is the case also here, as the IP type IV burst is
observed only in the STEREO-A view at source longitude E42. 

The F-H emission lane pair is clearly visible during 05:00\,--\,06:00 UT.
However, the emission between 04:00 UT and 05:00 UT, that looks to
belong to the F-lane, does not show clear H-lane emission
(Figure \ref{fig:wind-stereo-0410}, start of F-H emission is indicated
in both {\it Wind} and STEREO-A dynamic spectra). 
The observed emissions drift from 320~kHz at 04:00 UT to 280~kHz at
06:00 UT, corresponding to heliocentric heights from 21.0~R$_{\odot}$
to 23.1~R$_{\odot}$. However, the curvature of emission towards the
higher frequencies could be a visual effect, if the emission at
04:00\,--\,05:00 is not part of the F-H type II burst.

The F-lane source heights after 05:00 UT match quite well with the
heights of the CME front observed by STEREO-A, and the white-light
shock region could propagate along the same height-time
path, see Figure \ref{fig:heights-2012-stereo}. The CME leading front
heights could be estimated from the STEREO-A/HI1-A images at heights
larger than 15~R$_{\odot}$, but the white-light shock front was visible
only in the STEREO-A/COR2 images. 

\section{Summary and Conclusions}
\label{conclusions}

This study reports on two similar-looking, two-part interplanetary (IP)
radio type II burst events from 2003 and 2012. In these events, a diffuse
wide-band type II burst was followed by a type II burst that shows fundamental
and harmonic (F-H) emission lanes, both of which are slightly curved in their 
frequency-time evolution.

The change from a diffuse wide-band lane to narrow F-H bands is not
typical for IP type II bursts. Of the 25 wide-band type II events from
2001-2011 presented in \cite{pohjolainen2013}, only a few suggest
such evolution. The 2003 event was listed there as event
number 9, and most of the bursts occurred before STEREO observations.
As the newer 2012 event was observed from three different viewing angles,
we could here compare the observations from each spacecraft.

The associated halo-type CMEs in the 2003 and 2012 events showed similar
shapes and structures. In both events, the elongated CME fronts had heights
that match with the wide-band diffuse type II emission, if we consider
the whole emission lane and not just the lane center. If the diffuse
type II emission is plasma radiation, and not synchrotron emission as
it has sometimes been suggested, then only the atmospheric densities and
the source heights determine the emitting frequencies. The widening of
the emission band could correspond to a larger emission region, or it
could be explained by a larger density gradient at the radio source region.
For example, \cite{iwai2020} found a positive correlation between the
type II bandwidth and SEP peak flux, supporting the idea that if particle
acceleration occurs on a larger spatial scale, more SEPs will be generated. 

In these two events, the diffuse type II burst weakens and disappears
from the spectrum before or near the time when the F-H type II burst is
formed. In the statistical study by \cite{pohjolainen2013}, they found
that in most cases the diffuse emission continues much longer,
and it can be followed to much lower frequencies. Hence, it could be
possible that the formation of the F-H type II burst is somehow connected
to the early ending of the diffuse radio emission. For example, the
wide leading front of the CME could break up, causing the shock region
to move down on the CME side, to a more narrow region.  

In both events, the F-H type II burst heights are much lower than the
highest observed CME front heights. A simple explanation would be that a
separate flank shock is formed at lower heights, due to the lateral CME
expansion and collision with a streamer. In the 2003 event a flank shock
looks probable, as we see a bending streamer near the heights of the
F-H type II burst. This could also explain the curvature of the F-H
emission lanes: at the CME flanks the atmospheric densities are first
lower, but when the shock meets increased streamer densities the emission
frequency increases. When the shock region moves forward, the shock
propagates back into decreasing densities.

In the 2012 event there are emission structures present after the ending
of the diffuse radio burst and before the appearance of the F-H burst,
which do not look to belong to either of the bursts. In this event the
white-light imaging observations do not show any clear indication of
interaction with streamers. 

In the earlier analysis of the 2012 event, \cite{shen2013} suggested
that the event consisted of two separate CMEs. In another study,
\cite{grechnev2024} identified two wave-like disturbances that later
merged into a halo-shock. Therefore they fitted the observed radio
emissions with three separate type II bursts (Figure 19 in their article).
Basically, their 1F-H is a continuation of the metric type II burst,
their 3F corresponds to our diffuse type II lane, and their 2F-H is located
in between them. \cite{grechnev2024} also suggested that at distances
exceeding ten solar radii the piston shock was transformed into a bow
shock.


Based on our current analysis and the earlier studies on the 2003 and
2012 events, the hypothesis for the two-part interplanetary type II
bursts are:

\begin{itemize}
\item A bow-shock is formed along the leading front of a CME, creating
a wide-band type II burst. The bow shock dies out when the front separates
into multiple loops. At heights $\approx$\,20~R$_{\odot}$, a CME flank shock
is formed, with interaction with a nearby streamer. The F-H type II emission
from the flank shock reflects the varying densities within the streamer
region.
\item Alternatively, together with the bow-shock that creates the
wide-band type II burst, a second global shock is formed at the same time.
Due to plasma and possible viewing angle effects, the F-H type II shock
becomes visible only later, at heights $\approx$\,20~R$_{\odot}$.
\item Third possible scenario is that the eruption creates two CME fronts
that propagate to slightly different directions, with different speeds.
The fast and wide CME front creates the diffuse wide-band type II burst,
and the slower CME creates the F-H type II burst later on. 
\end{itemize}

In the 2003 event, a bow shock with a later flank shock and interaction
with a streamer looks possible. We found no evidence of more than one
CME or other white-light shocks, but as the existing observations were
limited to one field-of view, with gaps in the coronagraph imaging, this
result can still be questioned. 

In the 2012 event, the wide-band diffuse type II emission was observed
only in the STEREO-A view. The CME height was, however, lower in the
STEREO-A view than in the SOHO/LASCO view. This height difference
could simply be due to the CME front being directed more towards
STEREO-A, and LASCO would observe the front better from a side angle.  
The existence of IP type IV radio emission, but only in the STEREO-A
view, also indicates a rising CME structure that is moving towards
STEREO-A. The diffuse radio type II burst emission agrees with a
propagating CME bow shock, in heights and speed, and an in-situ shock
was observed to arrive to STEREO-A in a time-frame comparable to the
CME speed. 

The STEREO-A coronagraph images also show a white-light shock region
located on the north-eastern side of the CME. This suggests a simultaneous
existence of a CME bow shock at LASCO heights, and a second, global
shock at STEREO-A heights. Comparison between the STEREO-A shock
heights and the later-appearing F-H type II burst heights suggest 
that the F-H type II burst could have been created by this shock.
The propagation speed of the shock structure was approximately
1100~km~s$^{-1}$. With deceleration, it could match with the shock arrival
near Earth 72 hours later. However, as the tracking of the white-light
shock region can be done only up to the COR2 coronagraph height limit of
15~R$_{\odot}$, we do not know what happens to the shock after that.  

So these questions remains open: are there two CMEs that create two
propagating shocks in the 2012 event, that move at different speeds to
slightly different directions, or is there just one CME that creates
two separate shocks? A fast and wide CME, directed toward the observing
instrument, would create a CME bow shock that shows up as a diffuse
wide-band type II burst. For the later-appearing F-H type II burst we
cannot rule out the possibility of a second CME, although the
one-CME and two-shocks scenario does have observational support. 

\section{Discussion}
\label{discussion}

Repeated events and similarities in emission structures have always
been of interest, especially for prediction purposes. 
Homologous CMEs that eject similar structures repeatedly from the same
source region have been reported by \cite{bian2022}.
Then, homologous solar flares present features that are most likely due
to similarities in the magnetic field configurations and particle
acceleration processes, see for example \cite{romano2018}, and references
therein. Particle beams and propagating shock waves create plasma emission
via nonlinear wave-wave interactions, but the emission contains also
various radio wave propagation effects in the inhomogeneous and turbulent
corona \citep{kontar2019}, that need to be taken into account. 

The trapped particles, indicated by the existence of a radio
type IV burst, could also affect the detected SEP populations. Recently,
\cite{jebaraj2024} presented the first-ever direct measurements of
synchrotron-emitting heliospheric traveling shocks, intercepted by
the Parker Solar Probe during its close encounters. They found that
strong quasi-parallel shocks are better emitters of radiation than
quasi-perpendicular shocks, due to the efficient acceleration of
ultra-relativistic electrons. Hence, it is of interest whether these
shocks show up as the diffuse wide-band type II burst, or if the
efficient acceleration is associated with the later, F-H part of a
type II burst event.

In the analysed 2012 event, the F-H type II burst was observed by STEREO-A
and {\it Wind}, in contrast to the diffuse type II and type IV bursts that
were observed only by STEREO-A. Moreover, the type IV burst ended near
the time when the F-H burst appeared. This may suggest something similar
to the cases presented in \cite{pohjolainen2020}, where a lower-located
type IV burst source is most visible when there is no shock front in
between the type IV source and the observer. A shock region producing
a type II burst could then act as an absorber, not letting the
higher-frequency type IV emission come through.   

Several studies that have used radio triangulation and 3D reconstruction
techniques have concluded that most of the type II burst sources are
located at the flanks of the CMEs \citep{magdalenic2014,hu2016,krupar2016}.  
Interaction with a streamer has often been suggested to be causing the
type II burst. In addition, \cite{jebaraj2020} found that one CME could be
the driver of two separate type II bursts, but the bursts appeared at
different parts of the CME, at different flanks. 

As we do not have radio imaging observations at DH wavelengths, the actual
radio source locations cannot be observed directly. The radio
triangulation technique and direction-finding observations have been used
as an indirect imaging tool in some cases, see \cite{jebaraj2021} and
references therein, but a more systematic and simple tool would be
needed, to explain the still-unknown features and physical processes
that occur associated with IP type II solar radio bursts.


\begin{acks}
Special thanks go to the people who have built and maintained the various
solar data archives, who contributed in creating and updating the solar
event catalogues, and thus enabled easy access to various data products.   
The CDAW CME catalogue is generated and maintained at the CDAW Data Center
by NASA and the Catholic University of America in cooperation with the Naval
Research Laboratory. SOHO is a project of international cooperation between
ESA and NASA. STEREO is part of the NASA Solar Terrestrial Probes (STP)
Program.
\end{acks}

\begin{ethics}
\begin{conflict}
The author declares there is no conflict of interests. 
\end{conflict}
\end{ethics}


\end{document}